\documentclass[pra,twocolumn,floatfix]{revtex4}
\usepackage{amsmath,amsfonts}
\usepackage{graphicx}
\usepackage{dcolumn}
\usepackage{color}
\usepackage{nicefrac}
\newcommand{\bfr}{\mathbf{r}}

\begin{document}

\title{Accurate and efficient approximation to the optimized
effective potential for exchange}

\author{Ilya G. Ryabinkin}
\author{Alexei A. Kananenka}
\author{Viktor N. Staroverov}
\affiliation{Department of Chemistry, The University of Western Ontario,
London, Ontario N6A 5B7, Canada}

\date{\today}

\begin{abstract}

We devise an efficient practical method for computing the Kohn--Sham
exchange-correlation potential corresponding to a Hartree--Fock
electron density. This potential is almost indistinguishable from
the exact-exchange optimized effective potential (OEP) and, when
used as an approximation to the OEP, is vastly better than all
existing models. Using our method one can obtain unambiguous, nearly
exact OEPs for any finite one-electron basis set at the same low cost
as the Krieger--Li--Iafrate and Becke--Johnson potentials. For all
practical purposes, this solves the long-standing problem of black-box
construction of OEPs in exact-exchange calculations.

\end{abstract}

\maketitle

The purpose of this Letter is to suggest an essentially exact, robust,
practical method for constructing the optimized effective potential
(OEP)~\cite{Grabo:2000/Anisimov/203} of the exact-exchange Kohn--Sham
scheme. OEPs naturally arise in the theory of orbital-dependent
functionals~\cite{Kummel:2008/RMP/3}---one of the most promising
modern density-functional techniques---and are of significant
practical interest because they afford qualitatively better
description of molecular properties than local and semilocal
approximations~\cite{Grabo:2000/Anisimov/203,Kummel:2008/RMP/3}.

The exchange-only OEP is defined~\cite{Sharp:1953/PR/317} as the
multiplicative potential $v_\text{X}^\text{OEP}(\bfr)$ that minimizes
the Hartree--Fock (HF) total energy expression within the Kohn--Sham
scheme. Equivalently~\cite{Sahni:1982/PRB/4371}, the OEP is the
functional derivative $v_\text{X}^\text{OEP}(\bfr)\equiv\delta
E_\text{X}^\text{exact}/\delta\rho(\bfr)$, where
$E_\text{X}^\text{exact}$ is the HF exchange energy
expression written in terms of Kohn--Sham orbitals
(an implicit density functional) and $\rho(\bfr)$ is the
electron density. To obtain $v_\text{X}^\text{OEP}(\bfr)$ in
a formally correct manner, one has to solve the OEP integral
equation~\cite{Grabo:2000/Anisimov/203}. Unfortunately, every
attempt to do this runs into severe numerical difficulties
because the problem is ill-posed~\cite{Hirata:2001/JCP/1635}
and has infinitely many solutions in finite basis
sets~\cite{Hirata:2001/JCP/1635,Staroverov:2006/JCP/141103}. Recent
advances in OEP methods~\cite{Kummel:2003/PRL/043004,%
Hesselmann:2007/JCP/054102,Gorling:2008/JCP/104104,%
Heaton-Burgess:2007/PRL/256401,Kollmar:2007/JCP/114104,%
Kollmar:2008/JCP/064101,Jacob:2011/JCP/244102,Gidopoulos:2012/PRA/052508}
have alleviated some of these difficulties but, even today, flawless
OEPs can be obtained only case by case, with painstaking effort.

In the absence of an efficient OEP solver, various approximations to
the OEP have long been used as pragmatic alternatives. These include
the Krieger--Li--Iafrate (KLI)~\cite{Krieger:1992/PRA/101}, localized
Hartree--Fock (LHF)~\cite{DellaSala:2001/JCP/5718}, and related
approximations~\cite{Gruning:2002/JCP/6435,Holas:2005/PRA/032504,%
Staroverov:2006/JCP/081104,Izmaylov:2007/JCP/084107},
as well as model potentials for exact
exchange~\cite{Leeuwen:1996/TCC/107,Gritsenko:1999/CPL/199,%
Becke:2006/JCP/221101,Staroverov:2008/JCP/134103,Rasanen:2010/JCP/044112},
of which the Becke--Johnson (BJ)
approximation~\cite{Becke:2006/JCP/221101} is the most popular. The
LHF method is equivalent~\cite{Izmaylov:2007/JCP/084107}
to the common energy denominator approximation
(CEDA)~\cite{Gruning:2002/JCP/6435} and to the effective local
potential (ELP) scheme~\cite{Staroverov:2006/JCP/081104}.

In a parallel development, several workers
studied~\cite{Payne:1979/JCP/490,Holas:1993/PRA/2708,%
Gorling:1995/PRA/4501,Holas:1996/TCC/57} the HF
method as a density-functional problem and occasionally
observed~\cite{Chen:1994/PMB/1001,Filippi:1996/PRA/4810} that
Kohn--Sham exchange-correlation potentials corresponding to HF
electron densities (HFXC potentials for short) were very close to
OEPs. However, this observation had little impact on the OEP
impasse because existing methods for determining exchange-correlation
potentials from densities (see, for instance,
Refs.~\onlinecite{Zhao:1994/PRA/2138,Leeuwen:1994/PRA/2421,%
Tozer:1996/JCP/9200,Wu:2003/JCP/2498,Ryabinkin:2012/JCP/164113})
face the same basis-set artifacts~\cite{Schipper:1997/TCA/16} and
numerical challenges~\cite{Bulat:2007/JCP/174101} as attempts to
solve the OEP equation.

In this work, we devise a practical, artifact-free procedure which
allows one to compute the HFXC potential efficiently for any atom or
molecule. Then we use our method to show, on a variety of systems, that
HFXC potentials are not just close but practically indistinguishable
from OEPs. The significance of our approach is that it has the same
reliability and computational cost as the KLI, LHF, and BJ schemes,
but its accuracy is vastly superior.

The proposed method originated with our observation that the quantity
$(\tau^\text{HF}-\tau)/\rho$, where $\tau$ and $\tau^\text{HF}$
are the Kohn--Sham and HF kinetic energy densities, reproduces that
part of atomic shell structure of exact-exchange potentials which
is missing in the KLI and LHF approximations. While searching for a
rigorous explanation, we realized that we were dealing with the HFXC
potential and arrived at the following argument.

Consider the HF description of a closed-shell $N$-electron
system. The exchange energy of this system is
\begin{equation}
 E_\text{X}^\text{HF} = -\frac{1}{4} \int d\bfr \int
 \frac{|\gamma^\text{HF}(\bfr,\bfr')|^2}{|\bfr - \bfr'|}\,d\bfr',
   \label{eq:EXX}
\end{equation}
where $\gamma^\text{HF}(\bfr,\bfr') = \sum_{i=1}^N
 \phi_i^\text{HF}(\bfr) \phi_i^{\text{HF}*}(\bfr')$
is the spinless reduced density matrix and $\phi_i^\text{HF}$
is the spatial part of the $i$th canonical HF spin-orbital. The
HF electron density is given by
$\rho^\text{HF}(\bfr) = \sum_{i=1}^N |\phi_i^\text{HF}(\bfr)|^2$.
The orbitals $\phi_i^\text{HF}$ are the lowest-eigenvalue solutions
of the HF equations
\begin{equation}
 \left[ -\dfrac{1}{2}\nabla^2 + v(\bfr) 
 + v_\text{H}(\bfr) + \hat{K} \right] \phi_i^\text{HF}(\bfr) 
 = \epsilon_i^\text{HF} \phi_i^\text{HF}(\bfr),   \label{eq:HF}
\end{equation}
where $v(\bfr)$ is the external potential (e.g., the potential of
the nuclei), $v_\text{H}(\bfr)
=\int \rho^\text{HF}(\bfr')|\bfr-\bfr'|^{-1}\,d\bfr'$
is the Hartree (electrostatic) potential of $\rho^\text{HF}(\bfr)$,
and $\hat{K}$ is the Fock exchange operator defined by
\begin{equation}
 \hat{K}\phi_i^\text{HF}(\bfr) 
 = \frac{\delta E_\text{X}^\text{HF}}{\delta\phi_i^{\text{HF}*}(\bfr)}
 = -\frac{1}{2} \int \frac{\gamma^\text{HF}(\bfr,\bfr')}{|\bfr - \bfr'|}
  \phi_i^\text{HF}(\bfr')\,d\bfr'.   \label{eq:k-def}
\end{equation}

Let us multiply Eq.~\eqref{eq:HF} by $\phi_i^{\text{HF}*}$, sum over
$i$ from 1 to $N$, and divide through by $\rho^\text{HF}$. The
result is
\begin{equation}
 \frac{\tau_L^\text{HF}}{\rho^\text{HF}} 
 + v + v_\text{H} + v_\text{S}^\text{HF}
 = \frac{1}{\rho^\text{HF}} \sum_{i=1}^N
 \epsilon_i^\text{HF} |\phi_i^\text{HF}|^2,
  \label{eq:HF-inv-interm}
\end{equation}
where $\tau_L^\text{HF}(\bfr) = -\frac{1}{2} \sum_{i=1}^N
\phi_i^\text{HF*}(\bfr)\nabla^2\phi_i^\text{HF}(\bfr)$
is the Laplacian form of the HF kinetic energy density and
\begin{equation}
 v_\text{S}^\text{HF}(\bfr)
 = -\frac{1}{2\rho^\text{HF}(\bfr)}
  \int \frac{|\gamma^\text{HF}(\bfr,\bfr')|^2}{|\bfr-\bfr'|}\,d\bfr'.
  \label{eq:Slater-def}
\end{equation}
is the Slater potential (the orbital-averaged $\hat{K}$
operator)~\cite{Slater:1951/PR/385} built from the HF orbitals. 
The quantity on the right-hand side of Eq.~\eqref{eq:HF-inv-interm}
is known as the HF average local ionization
energy~\cite{Bulat:2009/JPCA/1384},
\begin{equation}
 \bar{I}^\text{HF}(\bfr)
 = \frac{1}{\rho^\text{HF}(\bfr)} \sum_{i=1}^N
 \epsilon_i^\text{HF} |\phi_i^\text{HF}(\bfr)|^2.  \label{eq:I-HF}
\end{equation}
Note that $\tau_L^\text{HF} = \tau^\text{HF}
-\frac{1}{4} \nabla^2\rho^\text{HF}$, where
\begin{equation}
 \tau^\text{HF}(\bfr) = \frac{1}{2}\sum_{i=1}^N
  |\nabla\phi_i^\text{HF}(\bfr)|^2   \label{eq:tau-HF}
\end{equation}
is the positive-definite form of the HF kinetic energy
density. In practical calculations, it is much better
to deal with $\tau^\text{HF}$ than with $\tau_L^\text{HF}$
because the former is always finite, whereas the latter
becomes infinite at the nuclei.
With these definitions we rewrite Eq.~\eqref{eq:HF-inv-interm} as
\begin{equation}
 \frac{\tau^\text{HF}}{\rho^\text{HF}} 
 - \frac{1}{4} \frac{\nabla^2\rho^\text{HF}}{\rho^\text{HF}}
 + v + v_\text{H} + v_\text{S}^\text{HF}
 = \bar{I}^\text{HF}.  \label{eq:HF-inv}
\end{equation}

Now, let us pose the following problem: Find the multiplicative
exchange-correlation potential of the Kohn--Sham scheme which
generates the same electron density as the HF method.
This HFXC potential, $v_\text{XC}^\text{HF}(\bfr)$, is defined by the
Kohn--Sham equations
\begin{equation}
 \left[ -\dfrac{1}{2}\nabla^2 + v(\bfr)
 + v_\text{H}(\bfr) + v_\text{XC}^\text{HF}(\bfr) \right]
  \phi_i(\bfr) = \epsilon_i \phi_i(\bfr),   \label{eq:KS}
\end{equation}
where $v$ and $v_\text{H}$ are the same as in Eq.~\eqref{eq:HF}
and the eigenfunctions $\phi_i$ are such that $\rho(\bfr)\equiv
\sum_{i=1}^N |\phi_i(\bfr)|^2 = \rho^\text{HF}(\bfr)$.  An important point
here is that the equality $\rho=\rho^\text{HF}$ does \textit{not}
imply that $\phi_i=\phi_i^\text{HF}$. In fact, the canonical
orbitals $\phi_i$ and $\phi_i^\text{HF}$ are known to be slightly
different~\cite{Gorling:1995/PRA/4501}.

To find $v_\text{XC}^\text{HF}(\bfr)$, we perform the same
manipulations on Eq.~\eqref{eq:KS} that led from Eq.~\eqref{eq:HF}
to Eq.~\eqref{eq:HF-inv} and arrive at
\begin{equation}
 \frac{\tau}{\rho} -\frac{1}{4}\frac{\nabla^2\rho}{\rho}
  + v + v_\text{H} + v_\text{XC}^\text{HF} = \bar{I},
  \label{eq:KS-inv}
\end{equation}
where $\tau(\bfr) = \frac{1}{2} \sum_{i=1}^N |\nabla\phi_i(\bfr)|^2$
is the positive-definite Kohn--Sham kinetic energy density, and
\begin{equation}
 \bar{I}(\bfr) = \frac{1}{\rho(\bfr)}
  \sum_{i=1}^N \epsilon_i |\phi_i(\bfr)|^2  \label{eq:I-KS}
\end{equation}
is the Kohn--Sham average local ionization energy.
Finally, we subtract Eq.~\eqref{eq:HF-inv}
from \eqref{eq:KS-inv} and write
\begin{align} 
 v_\text{XC}^\text{HF}(\bfr) 
 & = v_\text{S}^\text{HF}(\bfr)
 + \bar{I}(\bfr) - \bar{I}^\text{HF}(\bfr)
 + \frac{\tau^\text{HF}(\bfr)}{\rho^\text{HF}(\bfr)}
 - \frac{\tau(\bfr)}{\rho(\bfr)},  \label{eq:v-final}
\end{align}
where $\rho=\rho^\text{HF}$, but $\tau\neq\tau^\text{HF}$
and $\bar{I}\neq\bar{I}^\text{HF}$.

Equation~\eqref{eq:v-final} is the key result of this
work. It gives the HFXC potential exactly (in a complete
basis). Analogous but less practical expressions for
$v_\text{XC}^\text{HF}$ were presented earlier in
Refs.~\onlinecite{Holas:1997/PRB/1295,Nagy:1997/PRA/3465,%
Miao:2000/PMB/409}.

We propose to treat Eq.~\eqref{eq:v-final} as the definition of
a model Kohn--Sham potential for exact exchange. To turn
this definition into a practical method we observe that $\bar{I}$
and $\tau$ are determined by $v_\text{XC}^\text{HF}$ and hence are
initially unknown. Therefore, Eq.~\eqref{eq:v-final} has to be solved
iteratively. The algorithm we suggest is as follows.

\begin{enumerate}
\setlength{\itemsep}{0.5ex}
\setlength{\parskip}{0.0ex}
\setlength{\parsep}{0.0ex}

\item Perform an HF calculation on the system of interest
and construct $\rho^\text{HF}$, $v_\text{S}^\text{HF}$,
$\tau^\text{HF}$, and $\bar{I}^\text{HF}$.

\item Choose an initial guess for the occupied
Kohn--Sham orbitals $\{\phi_i\}$ and their eigenvalues
$\{\epsilon_i\}$ (e.g., HF orbitals and orbital energies).

\item Shift all $\epsilon_i$ simultaneously to satisfy the condition
$\epsilon_N=\epsilon_N^\text{HF}$. This is needed to ensure that
$v_\text{XC}^\text{HF}$ retains the correct $-1/r$ asymptotic behavior
of $v_\text{S}^\text{HF}$.

\item Construct $v_\text{XC}^\text{HF}$ by substituting the current
$\{\phi_i\}$ and $\{\epsilon_i\}$ into Eq.~\eqref{eq:v-final}.
To facilitate convergence, we found it essential to compute the
terms $\bar{I}$ and $\tau/\rho$ using the density $\rho=\sum_{i=1}^N
|\phi_i|^2$ rather than $\rho^\text{HF}$.

\item Solve the Kohn--Sham equations \eqref{eq:KS} using the current
$v_\text{XC}^\text{HF}$. This gives a new set of $\{\phi_i\}$
and $\{\epsilon_i\}$.

\item Return to Step 3. Iterate until
$v_\text{XC}^\text{HF}$ is self-consistent, i.e., until
$\{\phi_i\}$ and $\{\epsilon_i\}$ on input and output agree
within a desired threshold.

\end{enumerate}

For spin-polarized systems, there will be two HFXC potentials
(spin-up and spin-down) and hence two sets of all quantities except
$v$ and $v_\text{H}$. The entire scheme described above was
implemented in \textsc{gaussian 09}~\cite{G09-short}.

The most computationally intensive step in the HFXC approach,
as in the KLI, LHF, BJ, and related approximations, is the
construction of the Slater potential. It helps that in our
method the Slater potential has to be computed only once (at
the start of iterations). To eliminate every possible source of
errors unrelated to the HFXC approximation, here we constructed
$v_\text{S}^\text{HF}(\bfr)$ by using Eq.~\eqref{eq:Slater-def}. For
routine applications, we recommend resolution-of-the-identity
techniques or the method of Ref.~\onlinecite{Kananenka:2013}.

To assess the quality of HFXC potentials produced by
our method we compared them to the exact (numerical)
OEPs, some of the best existing OEP approximations (KLI,
ELP=LHF=CEDA, and BJ), and finite-basis-set OEPs obtained by
the Wu--Yang OEP (WY-OEP) method~\cite{Wu:2003/JTCC/627}. The
OEP and KLI results were taken from the work of Engel and
coworkers~\cite{Engel:1993/PRA/2800,Engel:1999/JCC/31,Engel:2013/private}
(for spherical atoms) and from Makmal \textit{et
al.}~\cite{Makmal:2009/JCTC/1731} (for molecules); these
are exact fully numerical solutions of the OEP and KLI
equations.  The BJ, ELP, and WY-OEP results were obtained
earlier by one of the authors~\cite{Gaiduk:2008/JCP/204101}. To
simulate the basis-set limit in the HFXC, BJ, ELP, and WY-OEP
calculations we employed the large universal Gaussian basis set
(UGBS) of Ref.~\onlinecite{Castro:1998/JCP/5225} for atoms and
UGBS1P (UGBS augmented with one set of polarization
functions for each exponent) for molecules. The accuracy of the UGBS
is such that total atomic HF energies computed in this basis are
converged to 7 significant figures with respect to the basis-set
limit~\cite{Castro:1998/JCP/5225}.

\begin{figure}
\centering
\includegraphics[width=0.48\textwidth]{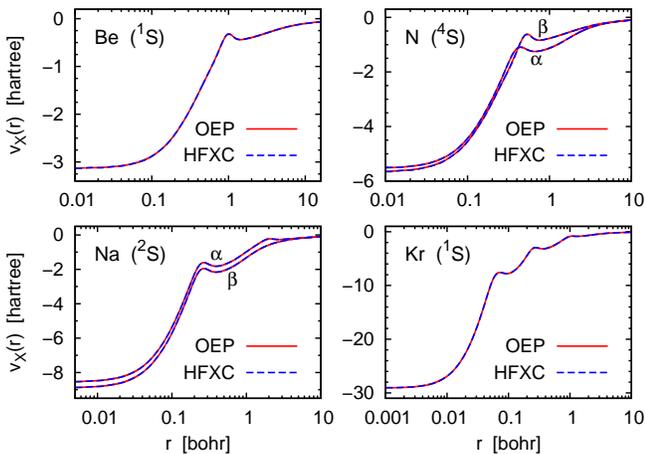}
\caption{OEPs and HFXC potentials are visually indistinguishable.
The same excellent agreement was observed for all atoms
where comparison with OEPs was made.}
\label{fig:1}
\end{figure}

\begin{figure}
\centering
\includegraphics[width=0.48\textwidth]{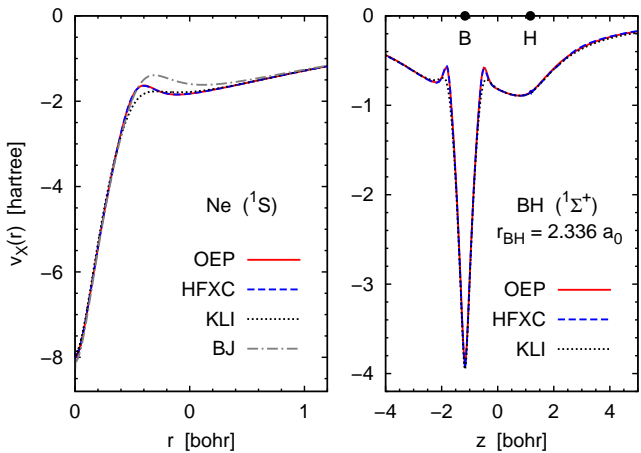}
\caption{HFXC potentials are perfect representations of
OEPs, unlike KLI and BJ potentials.
The potentials for the BH molecule are shown along the internuclear axis.}
\label{fig:2}
\end{figure}

In all cases where the UGBS (UGBS1P) was used, we found that
HFXC potentials are virtually indistinguishable from exact
OEPs (Figs.~\ref{fig:1} and~\ref{fig:2}) and are dramatically
better as approximations to OEPs than the KLI and BJ models
(Fig.~\ref{fig:2}). Note that the performance of the LHF approximation
is very similar to that of the KLI~\cite{DellaSala:2001/JCP/5718}
scheme, so the LHF or ELP or CEDA curves (not shown in
Fig.~\ref{fig:2}) would be almost superimposed with the KLI
potentials. The excellent agreement between HFXC potentials and exact
OEPs suggests that the `correlation' part of an HFXC potential is
negligibly small.

For quantitative comparison, we took the self-consistent Kohn--Sham
orbitals generated by HFXC and other potentials and calculated
the conventional total exchange-only energy, $E_\text{conv}$,
which defined as the HF total energy expression in terms of
Kohn--Sham orbitals. Table~\ref{tab:1} shows that the KLI, ELP, and
BJ potentials produce $E_\text{conv}$ values noticeably above the
exact OEP energies. By contrast, conventional energies obtained from
HFXC potentials are within 0.1 m$E_\text{h}$ of the OEP benchmarks
for most atoms---closer than $E_\text{conv}$ values from WY-OEPs.

\begin{table*}
\caption{Total ground-state energies of 12 representative atoms
obtained with various exchange potentials.
The numerical OEP and KLI results are from Refs.~\onlinecite{Engel:1993/PRA/2800,%
Engel:1999/JCC/31,Engel:2013/private}. All other values
were computed using a large Gaussian basis set (UGBS).}
\begin{ruledtabular}
\begin{tabular}{l*{11}d}
 & \multicolumn{1}{c}{$E_\text{OEP}$}
 & \multicolumn{5}{c}{$E_\text{conv}-E_\text{OEP}$ (units of m$E_\text{h}$)}
 & \multicolumn{5}{c}{$E_\text{vir}-E_\text{conv}$ (units of m$E_\text{h}$)}
  \\ \cline{3-7} \cline{8-12}
 Atom & \multicolumn{1}{c}{(units of $E_\text{h}$)} &
 \multicolumn{1}{c}{KLI} & \multicolumn{1}{c}{ELP\footnotemark[1]}
  & \multicolumn{1}{c}{BJ} & \multicolumn{1}{c}{WY-OEP} 
 & \multicolumn{1}{c}{HFXC} &
 \multicolumn{1}{c}{KLI} & \multicolumn{1}{c}{ELP\footnotemark[1]} 
 & \multicolumn{1}{c}{BJ} & \multicolumn{1}{c}{WY-OEP} 
 & \multicolumn{1}{c}{HFXC} \\ \hline
Li &    -7.43250 &  0.06 &  0.08 &  1.20 &  0.00 &  0.00 
        & -5.28 &  4.61 & 50.45 & -0.01 & -0.04 \\
Be &   -14.57243 &  0.15 &  0.15 &  0.75 &  0.01 & -0.01 
        & -21.20 & 13.85 & 31.68 &  0.07 & -0.10 \\
N  &   -54.40340 &  0.36 &  0.34 &  4.14 &  0.01 &  0.00 
        & 24.74 & 78.47 & 250.56 & -0.04 & -0.21 \\
Ne &  -128.54541 &  0.58 &  0.57 &  9.59 &  0.02 &  0.01 
        & 155.62 & 197.51 & 781.68 & -0.05 & -0.14 \\
Na &  -161.85664 &  0.73 &  0.73 &  7.71 &  0.02 &  0.00 
        & 183.10 & 231.84 & 805.90 &  0.22 & -0.28 \\
Mg &  -199.61158 &  0.87 &  0.87 &  5.86 &  0.02 &  0.00 
        & 182.26 & 267.70 & 799.35 & -0.64 & -0.26 \\
P  &  -340.71500 &  1.28 &  1.28 &  3.99 &  0.02 & -0.03 
        & 144.86 & 376.08 & 904.12 &  1.21 & -1.84 \\
Ar &  -526.81222 &  1.74 &  1.83 &  3.36 &  0.09 & -0.07 
        & 124.68 & 512.68 & 1182.26 &  2.54 & -4.08 \\
Ca &  -676.75193 &  2.23 &  1.98 &  3.58 & -0.03 & -0.13 
        & 14.94 & 597.73 & 1126.91 & -2.49 & -5.86 \\
Zn & -1777.83436 &  3.65 &  3.05 & 10.40 & -0.07 & -0.07 
        & 1047.87 & 1238.61 & 2130.22 & -1.21 & -5.93 \\
Kr & -2752.04295 &  3.18 &  3.44 &  6.52 &  0.26 & -0.07 
        & 1468.11 & 1657.32 & 3128.48 &  7.85 & -7.43 \\
Cd & -5465.11441 &  6.04 &  5.58 &  6.46 &  0.92 & -0.26 
        & 1883.92 & 2374.14 & 3617.48 & -4.25 & -6.99 \\ [0.5ex]
 m.a.v.\footnotemark[2] & & 1.74 & 1.66 & 5.30 & 0.12 & 0.05 
     & 438.0 & 629.2 & 1234.1 & 1.76 & 2.76 \\
\end{tabular}
\end{ruledtabular}
\footnotetext[1]{The ELP method is equivalent
to the LHF and CEDA schemes with frozen HF orbitals.}
\footnotetext[2]{Mean absolute value.}
\label{tab:1}
\end{table*}

A more stringent quality test~\cite{Gaiduk:2008/JCP/204101}
for OEP approximations is the virial energy discrepancy,
$\Delta_\text{vir}=E_\text{vir}-E_\text{conv}$, where
$E_\text{vir}$ is the total energy with the exchange
contribution obtained by the Levy--Perdew virial
relation~\cite{Levy:1985/PRA/2010},
\begin{equation}
 E_\text{X}^\text{vir}
 = \int v_\text{X}(\bfr) \left[ 3\rho(\bfr)
  + \bfr\cdot\nabla\rho(\bfr) \right] d\bfr.    \label{eq:LP}
\end{equation}
For exact OEPs, $\Delta_\text{vir}=0$~\cite{Ou-Yang:1990/PRL/1036}.
Table~\ref{tab:1} shows that virial energy discrepancies for
HFXC potentials do not exceed a few m$E_\text{h}$, that is, are
three orders of magnitude smaller than for the LHF, ELP, and BJ
approximations---as small as for WY-OEPs. These discrepancies are
expected to be even smaller in the basis-set limit. (The numerical
OEPs have $\Delta_\text{vir}$ values of the order of a few $\mu
E_\text{h}$~\cite{Engel:2013/private}.)

\begin{figure}
\centering
\includegraphics[width=0.48\textwidth]{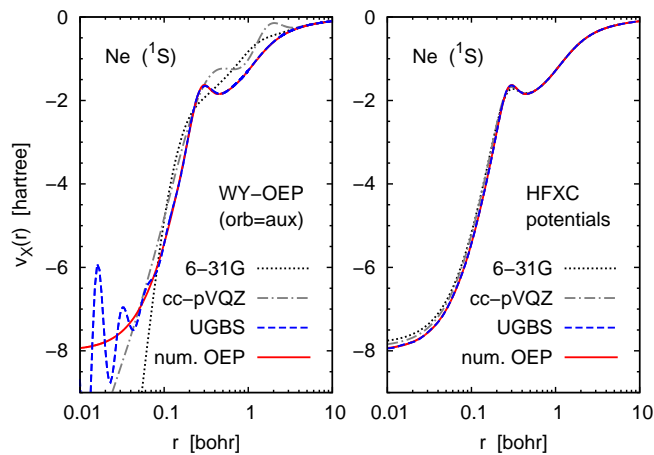}
\caption{WY-OEPs and HFXC potentials obtained with small (6-31G),
intermediate (cc-pVQZ), and large (UGBS) basis sets as 
approximations to the exact (numerical) OEP. The
HFXC/UGBS curve is right on top of the OEP.}
\label{fig:3}
\end{figure}

Recall that to solve the OEP integral equation by the WY method
one needs two sets of basis functions: a one-electron basis
for the orbitals and an auxiliary basis for the OEP. The two
sets must be ``balanced" with respect to each other; otherwise, the
resulting potential will be either suboptimal or highly
oscillatory~\cite{Staroverov:2006/JCP/141103,Hesselmann:2007/JCP/054102,%
Gorling:2008/JCP/104104,Heaton-Burgess:2007/PRL/256401}. By
employing the same large basis set in both roles one can
usually~\cite{Staroverov:2006/JCP/141103} obtain OEPs that are
smooth and correct everywhere except near the nucleus (the left
panel in Fig.~\ref{fig:3}). However, the single-basis trick does
not work for small and medium-sized one-electron basis sets such as
6-31G and cc-pVQZ, for which a suitable auxiliary basis can be found
only in an \textit{ad hoc} manner with considerable effort and some
arbitrariness~\cite{Hesselmann:2007/JCP/054102,Gorling:2008/JCP/104104,%
Heaton-Burgess:2007/PRL/256401}.
Such problems do not exist in our method, where we automatically
obtain a smooth HFXC potential for any one-electron basis
(the right panel in Fig.~\ref{fig:3}).  Since OEPs and HFXC
potentials are nearly identical in the basis-set limit, one can even
operationally define a finite-basis-set OEP (a fundamentally ambiguous
quantity~\cite{Staroverov:2006/JCP/141103}) as the corresponding
HFXC potential.

The reason the HFXC scheme is very robust is because the potential
$v_\text{XC}^\text{HF}$ is built up directly as a sum of commensurate,
well-behaved terms. Apart from being a tool for generating approximate
OEPs, the HFXC method can be used to determine Kohn--Sham potentials
from HF densities, provided that the HF and Kohn--Sham orbitals
are expanded in a complete (in practice, very large) basis or
represented on a dense grid. In Kohn--Sham calculations using a
\textit{finite} basis set, however, the potential given by
Eq.~\eqref{eq:v-final} reproduces the target $\rho^\text{HF}(\bfr)$
only approximately because Eq.~\eqref{eq:KS} and its finite-dimensional
matrix representation are not equivalent~\cite{footnote1}.

We can also identify the reason why HFXC potentials are
much closer to OEPs than KLI, LHF, and related approximations.
This happens because in our derivation we did \textit{not} assume
that $\phi_i=\phi_i^\text{HF}$ for all $i\leq N$. If, for the sake
of argument, we make this assumption in Eq.~\eqref{eq:v-final},
we immediately obtain a different potential,
\begin{equation}
 \tilde{v}_\text{XC}(\bfr) = v_\text{S}^\text{HF}(\bfr)
 + \frac{1}{\rho^\text{HF}(\bfr)}
  \sum_{i=1}^N (\epsilon_i - \epsilon_i^\text{HF}) |\phi_i^\text{HF}(\bfr)|^2,
  \label{eq:v-eps}
\end{equation}
which was introduced and discussed by Nagy~\cite{Nagy:1997/PRA/3465}
(with $\phi_i$ in place of $\phi_i^\text{HF}$) as an approximate
equivalent of the KLI potential. The difference between HF
and OEP orbitals may be small, but it gives rise to the crucial
$(\tau^\text{HF}-\tau)/\rho$ term responsible for the atomic shell
structure of the OEP. It follows that the KLI and LHF approximations
would be greatly improved simply by including this term.

In conclusion, we have shown (a) how to construct HFXC potentials
(i.e., model exchange-correlation potentials yielding HF densities in
the basis-set limit) at the computational cost of the KLI, LHF, and BJ
approximations; (b) that HFXC potentials are nearly exact approximations
to exchange-only OEPs, much better than the KLI, LHF, BJ, and related
models. The advantage of approximating OEPs with HFXC potentials is
that it the HFXC method completely avoids the OEP equation, and so is
free from numerical difficulties and basis-set artifacts that beset
OEP techniques.

HFXC potentials obtained in finite basis sets exhibit no spurious
oscillations and, for all intents and purposes, may be treated as
solutions of the OEP equation. In this sense, the HFXC scheme solves
the long-standing problem of unambiguous ``black-box" construction of
elusive finite-basis-set OEPs. We anticipate that our approach will
be widely embraced as a practical substitute for OEP methods and as a
superior alternative to existing model potentials for exact exchange.

Finally, we wish to remark that our approach can be generalized to
\textit{any} orbital-dependent exchange-correlation functional. One
simply needs to start with the corresponding energy expression
$E_\text{XC}[\{\phi_i\}]$ instead of $E_\text{X}^\text{HF}$
and modify appropriately all the steps in the derivation.
For $\tau$-dependent functionals and hybrids (mixtures of
exact exchange and semilocal approximations), this scheme is
expected to produce even more accurate approximations to $\delta
E_\text{XC}[\{\phi_i\}]/\delta\rho$ than for the exact-exchange
functional itself.

The authors thank Profs.~Eberhard Engel and Leeor Kronik for providing
the OEP and KLI benchmarks. I.G.R. is grateful to Dr.~Alex Gaiduk for help
with the \textsc{gaussian} code. This work was supported by the Natural
Sciences and Engineering Research Council of Canada (NSERC) through
the Discovery Grants Program.


\begin{thebibliography}{56}
\bibitem[{foo()}]{footnote1}
\bibinfo{note}{Problems such as ``find the multiplicative potential which
  reproduces the HF/6-31G density when the Kohn--Sham equations are solved in
  the 6-31G basis" are ill-posed. Using the method of
  Ref.~\onlinecite{Staroverov:2006/JCP/141103} it is easy to construct any
  number of potentials that satisfy the above requirement, but to
  decide which of these potentials is ``the true one" is fundamentally
  impossible~\cite{Harriman:1986/PRA/29}.}
\end{thebibliography}

\end{document}